\documentclass[preprint,showpacs,showkeys,preprintnumbers,amsmath,amssymb,PRL,endfloats,floatfix]{revtex4}

\usepackage{graphicx}% Include figure files
\usepackage{dcolumn}% Align table columns on decimal point
\usepackage{bm}% bold math

\begin{document}

\preprint{}

\title{Electrical detection of spin echoes for phosphorus donors in silicon}% Force line breaks with \\

\author{Hans Huebl}
\email[corresponding author ]{hans.huebl@unsw.edu.au}
\altaffiliation[ present address:~]{Australian Research Council
Centre of Excellence for Quantum Computer Technology, School of
Physics, University of New South Wales, Sydney NSW 2052, Australia}
\affiliation{Walter Schottky Institut, Technische Universit\"{a}t
M\"{u}nchen, Am Coulombwall 3, 85748 Garching, Germany}
\author{Felix Hoehne} \affiliation{Walter
Schottky Institut, Technische Universit\"{a}t M\"{u}nchen, Am Coulombwall 3,
85748 Garching, Germany}
\author{Benno Grolik} \affiliation{Walter
Schottky Institut, Technische Universit\"{a}t M\"{u}nchen, Am Coulombwall 3,
85748 Garching, Germany}
\author{Andre R. Stegner} \affiliation{Walter
Schottky Institut, Technische Universit\"{a}t M\"{u}nchen, Am Coulombwall 3,
85748 Garching, Germany}
\author{Martin Stutzmann} \affiliation{Walter
Schottky Institut, Technische Universit\"{a}t M\"{u}nchen, Am Coulombwall 3,
85748 Garching, Germany}
\author{Martin S. Brandt} \affiliation{Walter
Schottky Institut, Technische Universit\"{a}t M\"{u}nchen, Am Coulombwall 3,
85748 Garching, Germany}

\date{\today}% It is always \today, today,

\begin{abstract}
The electrical detection of spin echoes via echo tomography is used
to observe decoherence processes associated with the electrical
readout of the spin state of phosphorus donor electrons in silicon
near a SiO$_2$ interface. Using the Carr-Purcell pulse sequence, an
echo decay with a time constant of $1.7\pm0.2~\rm{\mu s}$ is
observed, in good agreement with theoretical modeling of the
interaction between donors and paramagnetic interface states.
Electrical spin echo tomography thus can be used to study the spin
dynamics in realistic spin qubit devices for quantum information
processing.
\end{abstract}

%Use showkeys class option if keyword
\maketitle A fundamental requirement for physical realizations of
quantum computers is that the decoherence time of the qubits
significantly exceeds typical gate operation times. As an example,
decoherence times of the electron spin of phosphorus donors of up to
60~ms have been reported in bulk isotopically pure $^{28}$Si
crystals \cite{tyryshkin03}. Since typical $\pi$ pulse lengths used
to manipulate such spins via magnetic resonance are of the order of
100~ns, donor spin states clearly satisfy the above requirement. The
effects of spin-spin interactions between neighboring P donors as
well as with the nuclear magnetic moments of the $^{29}$Si isotope,
which is  present with an abundance of 4.6\% in natural silicon, on
the decoherence have already been investigated in detail
\cite{chiba72, abe04}. However, the existing bulk measurements are
not directly relevant for quantum computing applications, since in
more realistic qubit devices such as proposed by Kane \cite{kane98},
the donors have to be in close proximity to a Si/SiO$_2$ interface,
which is required for the electrical isolation of gate electrodes
used to address the donor qubits \cite{kane98, martins04}.
Therefore, a coupling to paramagnetic interface states such as the
$\rm{P_{b0}}$ centers \cite{poindexter81, lenahan98} will provide an
additional source of decoherence \cite{schenkel06, desousa07}.
Stegner et al. \cite{stegner06} have actually used this
donor-interface state interaction for a purely electrical readout of
the donor spin states by pulsed electrically detected magnetic
resonance (pEDMR), which has the potential to overcome the
sensitivity limitations of conventional electron spin resonance
(ESR) down to the single spin detection limit \cite{mccamey06}. In
this paper, we demonstrate that pulsed EDMR experiments can also be
employed to investigate decoherence processes using a Carr-Purcell
pulse sequence in combination with echo tomography
\cite{childress06}. In Si/SiO$_2$ heterostructures with a natural
isotope composition, where the P donors are located within 15~nm of
the interface, we find that the Carr-Purcell echo at 6.5~K decays
with a time constant of $1.7\pm0.2~\rm{\mu s}$ and quantitatively
discuss this result taking into account decoherence and
recombination processes. In the framework of quantum information
processing, this result shows that the spin read-out via
P-$\rm{P_{b0}}$ spin pairs could be applied at least to fundamental
algorithms requiring up to $30\pi$ pulses. More importantly, this
work shows that with the help of advanced pulse sequences combined
with echo tomography valuable new information on the complex charge
carrier and spin dynamics can be obtained by EDMR.

\begin{figure*}[h]
\includegraphics[width=15cm]{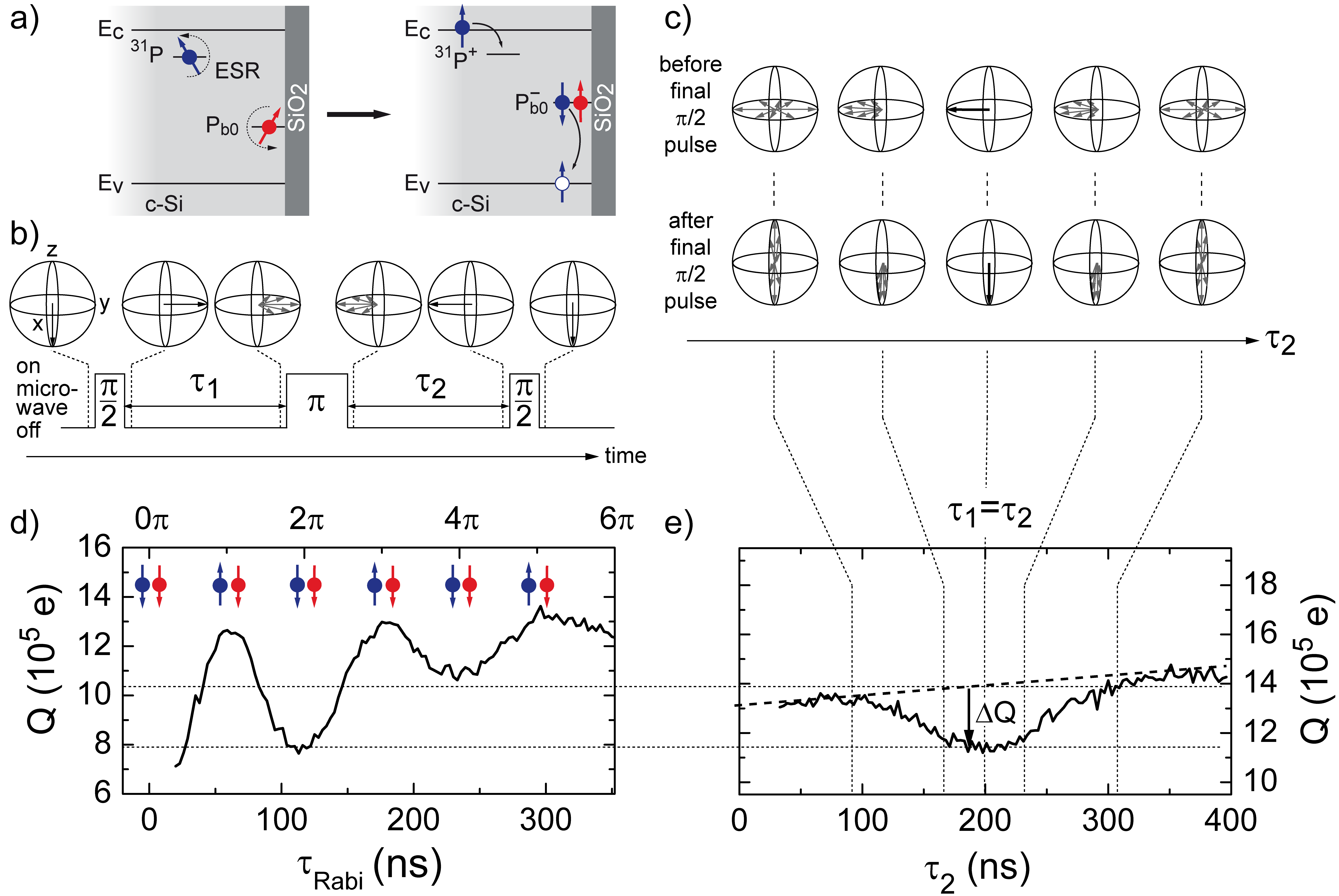}
\caption{a) Spin-dependent transition involving the electron spin of
a $^{31}$P donor and a paramagnetic state $\rm{P_{b0}}$ at the
interface between the crystalline Si (c-Si) and SiO$_2$. Since the
final state is a doubly occupied diamagnetic $\rm{P_{b0}^-}$, the
transition can only occur for initial P-$\rm{P_{b0}}$ spin pairs in
a singlet configuration. Electron spin resonance of either $^{31}$P
or $\rm{P_{b0}}$ influences the spin configuration, and results in a
resonant change of the transition rate which can be monitored by
charge transport through the test structure. b) Carr-Purcell echo
for an ensemble of identical spins plotted on a Bloch sphere. Before
the second $\pi/2$ pulse, an echo develops in the $x$-$y$-plane,
which is rotated to an eigenstate by the final ''echo tomography''
pulse, necessary for the electrical detection of the echo. The
effect of a change in the second evolution time $\tau_2$ with
respect to $\tau_1$ on the spin ensemble before and after the final
$\pi/2$ pulse is shown in more detail in c). d) Charge $Q$ obtained
from the integration of the current transients after microwave
pulses resonant with the $^{31}$P donor spin ensemble (Rabi
oscillations). For spin configurations of the P-$\rm{P_{b0}}$ spin
pair identical to the initial configuration (corresponding to
$2\pi$, $4\pi$, \ldots pulses), the transients and therefore $Q$ are
small. When the microwave pulse has brought the spin system out of
equilibrium, $Q$ is large ($\pi$, $3\pi$, \ldots pulses). e) The
total pulse length in a Carr-Purcell echo experiment including the
tomography pulse is $2\pi$. If $\tau_2=\tau_1$, the initial spin
configuration of the P-$\rm{P_{b0}}$ spin pair will be restored,
while for $\tau_2\neq\tau_1$, the spin system is out of equilibrium.
A comparison with the Rabi oscillation in d) predicts a decrease of
$Q$ under echo conditions $\tau_2\approx\tau_1$. Parts d) and e)
show the Rabi oscillation and the Carr-Purcell echo, respectivly, as
detected on the high-field resonance of the $^{31}$P donor electron
spin. The amplitude $\Delta Q$ of the echo agrees well with the
prediction. }\label{fig:schematics}
\end{figure*}

To detect the donor spin information electrically, we use a
spin-to-charge conversion method based on the Pauli exclusion
principle. The P donor and a nearby $\rm{P_{b0}}$ center form a
correlated spin pair as shown in Fig.~\ref{fig:schematics}~a). Since
a doubly occupied, negatively charged $\rm{P_{b0}^-}$ state as the
final state of the spin-to-charge conversion has a total spin $S=0$,
the transition of the electron from a P donor to a neutral
$\rm{P_{b0}}$ center is forbidden for triplet configurations of the
P-$\rm{P_{b0}}$ pairs, while for singlet P-$\rm{P_{b0}}$ pairs the
transition can occur, thus influencing the electronic transport
properties of the sample \cite{kaplan76}. Under spin resonance
conditions the relative orientation in the spin pair can be changed
coherently e.g.~from the triplet to the singlet configuration by a
$\pi$-pulse, resulting in a change of the rate of the spin-dependent
transition sketched in Fig.~\ref{fig:schematics}~a), which in turn
is observable as a resonant change of the conductivity
\cite{stegner06}.

Decoherence is usually studied with the help of a Carr-Purcell echo
experiment with a $\pi/2$-$\tau_1$-$\pi$-$\tau_2$-$echo$ pulse
sequence, where $\pi/2$ and $\pi$ denote the rotation angle of the
spin system induced by resonant microwave pulses and $\tau_{1}$ and
$\tau_{2}$ are the free evolution periods between the pulses
\cite{carr54}. Such a Carr-Purcell echo is shown in
Fig.~\ref{fig:schematics}~b) for an ensemble of identical spins
(e.g.~the spins of phosphorus donors) plotted in a Bloch sphere,
starting e.g.~with the spin ensemble in the ''down'' eigenstate
$|\downarrow\rangle$ \footnote{Since only the relative orientation
of the spin pairs is detected in EDMR, our arguments also apply to
spins in the ''up'' state $|\uparrow\rangle$ and therefore no
polarization of the spin ensemble is needed to observe echoes with
this technique.}. The microwave pulses are assumed to rotate the
spins around the $x$-axis of the Bloch sphere. The echo develops in
the $x$-$y$ plane of the Bloch sphere, giving rise to a pulse in the
transverse magnetization or coherence, which is easily detectable in
conventional ESR. However, the spin-to-charge conversion process
employed here is sensitive to the singlet/triplet symmetry of a spin
pair, which is not influenced by the orientation in the $x$-$y$
plane of the Bloch sphere of  one constituent spin if the other spin
is left in an eigenstate. A successful detection of such echoes via
charge transport therefore requires so-called echo tomography
\cite{childress06}, where after the second free evolution period
$\tau_2$ a final $\pi/2$-pulse rotates the spin system back into
singlet or triplet eigenstates of the pair. Again for an ensemble of
spins such as the P donor spins the effect of a final $\pi/2$ pulse
is shown in Fig.~\ref{fig:schematics}~c) on a Bloch sphere in more
detail. For $\tau_2 \ll \tau_1$ and $\tau_2 \gg \tau_1$, no echo has
developed in the $x$-$y$ plane so that after the final $\pi/2$
pulse, the spins of the ensemble point to all directions in the
$x$-$z$ plane of the Bloch sphere. Both, triplet and singlet
configurations will therefore be found in ensembles of
P-$\rm{P_{b0}}$ spin pairs. However, for $\tau_2=\tau_1$, an echo
has developed, so that after the final $\pi/2$ pulse, the spin
ensemble is in the original $|\downarrow\rangle$ eigenstate again.
If the $\rm{P_{b0}}$ partner in the spin state is also in the
$|\downarrow\rangle$ state, we find only the triplet configuration
$|\downarrow\downarrow\rangle$ for the P-$\rm{P_{b0}}$ spin pairs. A
comparison of the two cases with $\tau_2=\tau_1$ and
$\tau_2\neq\tau_1$ shows that via the application of the final
$\pi/2$ pulse, echoes can be formed also in the singlet/triplet
symmetry of spin pairs and therefore are accessible to purely
electrical detection.

In electrically detected magnetic resonance, until now only rotary
spin echoes have been reported \cite{boehme03b}, where the spin
system is continuously driven by the microwave and the echo develops
when the microwave-induced rotation is canceled out by an equally
long reversed rotation realized by a 180$^\circ$ phase shift in the
microwave. While these echoes in principle are also able to
quantitatively measure decoherence \cite{solomon59}, the use of
pulse sequences including free evolution periods together with echo
tomography is more versatile and in principle allows other
properties such as the exchange interaction in the P-$\rm{P_{b0}}$
spin pair or the superhyperfine interaction with $^{29}$Si to be
investigated by double electron electron resonance (DEER)
\cite{biehl75} and electron spin echo envelope modulation (ESEEM)
\cite{abe04} pulse techniques, respectively.

Due to RC timeconstants of the experimental system consisting of the
sample and the measurement electronics, Carr-Purcell echoes cannot
be observed in real time. Rather, current transients after the
application of the microwave pulse sequence are measured, where the
excited spin system relaxes back into the steady-state. Theory shows
that the amplitudes of the multi-exponential decay are proportional
to the deviation of the singlet/triplet configuration of the
spin-system at the end of the pulse sequence from the steady-state
configuration \cite{boehme03}. Therefore, measurements on long time
scales compatible with the detection system allow to monitor the
coherent manipulation of the spin system and its decoherence taking
place on much shorter timescales.

To increase the signal-to-noise ratio, the current transients are
integrated in a box-car type of analysis, yielding a characteristic
charge $Q$ as the primary result of the experiment. As shown in
Fig.~\ref{fig:schematics}~d) for the coherent manipulation of the P
spin by a microwave pulse of varying time $\tau_{Rabi}$, $Q$ is
small if the manipulation brings the spin system back into the
steady state. In contrast, the current transient and therefore $Q$
is large when the spin system has been brought out of equilibrium by
the microwave pulse. These electrically detected Rabi oscillations
have been discussed in detail by Stegner et al.~\cite{stegner06}.
Due to the Zeeman interaction, the spin state
$|\downarrow\downarrow\rangle$  is the energetically lowest state of
the P-$\rm{P_{b0}}$ spin pair and is therefore indicated as the
steady-state configuration in Fig.~\ref{fig:schematics}~d).

We are now in the position to predict the experimental signature of
the echoes discussed above. The pulse sequence
$\pi/2$-$\tau_1$-$\pi$-$\tau_2$-$\pi/2$ contains microwave pulses
with a total length of $2\pi$. Ideally, we therefore expect a value
of $Q$ after a Carr-Purcell echo sequence with $\tau_1=\tau_2$ equal
to the $Q$ found after a rotation by $2\pi$ in a Rabi-flop
experiment as shown in Fig.~\ref{fig:schematics}~e). For
$\tau_2\gg\tau_1$ and $\tau_2\ll\tau_1$, our discussion of
Fig.~\ref{fig:schematics}~c) above has shown that the spin pairs are
not in a steady-state configuration, and therefore a larger $Q$ is
expected as the result of such Carr-Purcell sequences, as also
sketched in Fig.~\ref{fig:schematics}~e).

The sample investigated here was grown by chemical vapor deposition
and consists of a 15~nm thick silicon layer doped with phosphorus at
$[{\rm{P}}]=10^{17}~\rm{cm^{-3}}$ covered with a native oxide on top
of a 500~nm thick nominally intrinsic buffer on a Si:B wafer
(30~$\Omega$cm) \cite{huebl06}. The measurements are performed at
9.765~GHz in a dielectric microwave resonator (Bruker ER 4118XMD5W1)
between 5~K and 15~K under illumination with the broad optical
spectrum of a tungsten lamp. The microwaves are generated by a
HP83640A synthesizer and amplified by a traveling wavetube
amplifier. For the conductivity measurements interdigit Cr/Au
contacts with a periodicity of 20~$\mu$m covering an active device
area of $2\times 2.25$~mm$^2$ are used, so that $\approx 10^{10}$ P
spins are probed. During the experiment, the sample is biased with
22~mV resulting in a current of $\approx 50~\rm{\mu A}$, which is
measured by a current amplifier. To record the current transients
the output of the amplifier (500~kHz bandwidth) is connected via a
passive Butterworth high-pass filter (7$^{th}$ order,
$f_{3dB}=400$~Hz) and a video amplifier (20~MHz bandwith) to a
digital storage oscilloscope. This allows to detect photocurrent
transients after the microwave pulse, which under these conditions
consist of a fast (characteristic time constant $t=3~\rm{\mu s}$)
rise of the conductivity, a slower ($t=11~\rm{\mu s}$) fall
overshooting the steady-state value followed by a very slow
($t=140~\rm{\mu s}$) rise back to the steady-state conductivity. An
unambiguous assignment of the experimental time constants to singlet
and triplet recombination times is not possible, in particular due
to the limited bandwidth of the current amplifier prohibiting the
observation of transients faster that 2-3~$\rm{\mu s}$. While a full
understanding of the time constants is certainly interesting in the
long run, it is not required for the experimental detection of
decoherence and the determination of the relevant time scales
performed here. To obtain a sufficient signal-to-noise level the
transients were accumulated with a repetition time of 350~$\mu$s.
Control experiments at slower repetition rates showed that the
results obtained were not influenced by this repetition time. The
box-car type integration was performed over the positive part of the
overall photocurrent transient from 2-22~$\rm{\mu}$s. Rabi
oscillations were used to determine the $\pi/2$-pulse time to 31~ns
at the specific microwave power employed corresponding to a of the
microwave $B_1$ field of 0.3~mT.

\begin{figure}[h]
\includegraphics[width=10cm]{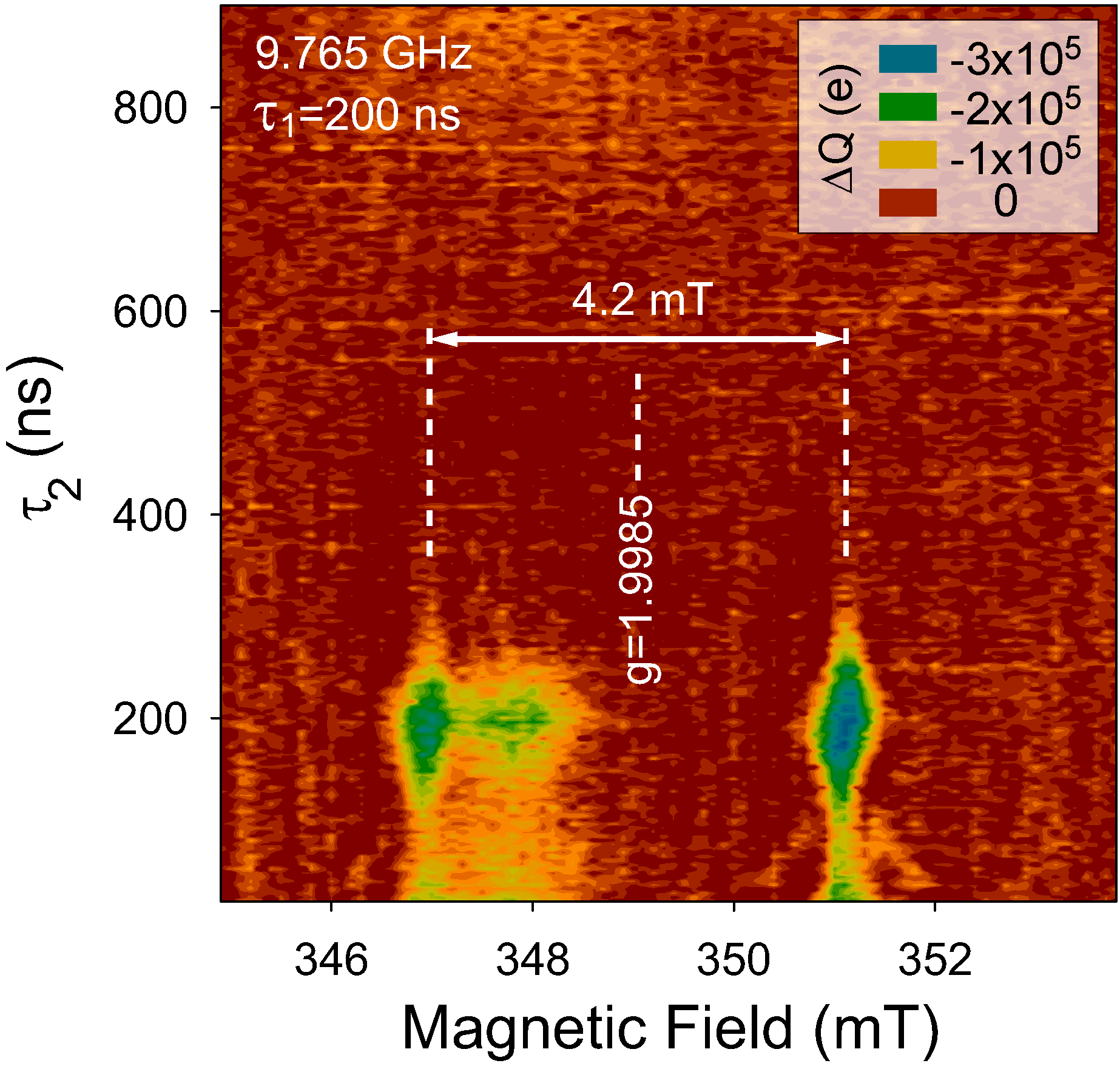}
\caption{Electrically detected Carr-Purcell echoes for a first free
evolution period of $\tau_1=200$~ns. A clear signature in the
integrated charge $\Delta Q$ appears at $\tau_1=\tau_2$ for all spin
resonance lines, while no further features are resolved up to
900~ns. The features at $\tau_2\lesssim 60$~ns are so-called Ramsey
fringes, which are observable for evolution periods $\tau_2$ below
the dephasing time of the system. A $\tau_{\pi/2}$ of 31~ns was
used, and the experiment was performed at
6.5~K.}\label{fig:CPecholong}
\end{figure}

Figure~\ref{fig:CPecholong} shows the Carr-Purcell echo tomography
data $\Delta Q$ obtained after a substraction of a constant
background from the Si:P/SiO$_2$ heterostructure. A first evolution
period $\tau_1=200$~ns was selected and $\tau_2$ as well as the
magnetic field $B_0$ were varied. As expected, a single echo with
$\Delta Q<0$ is found for $\tau_1=\tau_2=200$~ns for the 4.2~mT
hyperfine split P resonances with the central $g$-factor $g=1.9985$
at $B_0=351.2$~mT and $B_0=347.0$~mT and a broad P$_{\rm{b0}}$
resonance. In fact, two resonance lines at $g=2.008$
($B_0=347.4$~mT) and $g=2.004$ ($B_0=348.1$~mT) are expected for the
magnetic field oriented along the [110] axis of the Si sample
\cite{poindexter81}. Due to the high excitation power used for the
pulses and their large inhomogeneous linewidth these two resonances
are not resolved in Fig.~\ref{fig:CPecholong}, but rather appear as
a single feature at $B_0=347.9$~mT. The full-width half-maximum
temporal extent $\Delta \tau_{1/2}$ of the echo is correlated with
the inhomogeneous Gaussian linewidth $\Delta B_{1/2}$ via
$\Delta\tau_{1/2}=(2h)/(g\mu_B \Delta B_{1/2})$, where $g$ is the
$g$-factor, $\mu_B$ is the Bohr magneton and $h$ is Planck's
constant \cite{schweiger01echowidth}. For an inhomogeneous linewidth
of 0.4~mT for the high-field P resonance at $B_0=351.2$~mT, a
$\Delta\tau_{1/2}$ of 170~ns is expected, in good agreement with the
value of 130~ns determined from the experimental echo data $Q$ for
this resonance shown in more detail in Fig.~\ref{fig:schematics}~e)
\footnote{The larger linewidth of the P$_{\rm{b0}}$ resonances leads
to a narrower temporal extent of the echoes observed on these
resonances as expected.}. A quantitative comparison of $Q$ observed
during the echo (Fig.~\ref{fig:schematics}~e) to the $Q$ observed in
the Rabi oscillation (Fig.~\ref{fig:schematics}~d) demonstrates that
the echo amplitude $\Delta Q$ is indeed as large as expected from
the discussion above. However, the absolute values of $Q$ are higher
in the echo experiment and correspond to $Q$ found in the Rabi
oscillations at high $\tau_{Rabi}$. This is most probably caused by
the limited attenuation of the microwave switch in the ''off''
state, which leads to a weak but continuous disturbance of the spin
system also during the free evolution periods.

\begin{figure}[h]
\includegraphics[width=10cm]{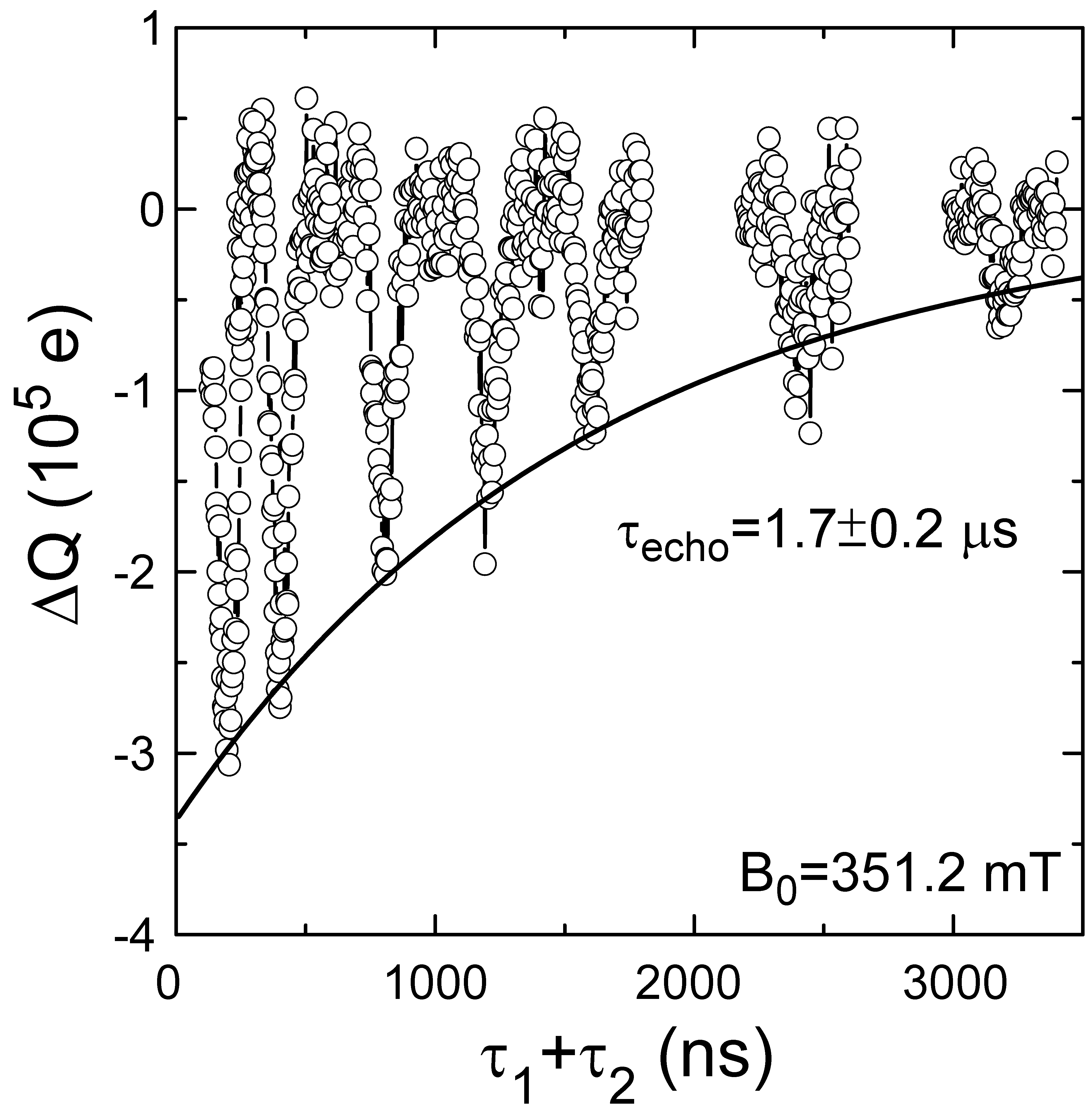}
\caption{Electrically-detected Carr-Purcell echoes observed on the
isolated high-field P resonance at $B_0=351.2$~mT for different
evolution times $\tau_1$. The solid line represents a
mono-exponential decay with a timeconstant of $1.7\pm0.2~\rm{\mu
s}$.}\label{fig:2dechoes}
\end{figure}

To determine the echo decay time in this system, the Carr-Purcell
echo sequence is measured as a function of $\tau_1$. In all cases,
the echo is observed at $\tau_1=\tau_2$, however the echo intensity
decreases for longer $\tau_1, \tau_2$. Figure~\ref{fig:2dechoes}
shows $\Delta Q$ as a function of the total free evolution period
$\tau_1+\tau_2$ again for the high-field P resonance at
$B_{0}=351.2$~mT obtained after substraction of a linear background
shown as a dashed line in Fig.~\ref{fig:schematics}~e). The decay of
the echo amplitude is described well with a mono-exponential decay
using a time constant $\tau_{echo}=1.7\pm0.2~\rm{\mu s}$ (solid line
in Fig.~\ref{fig:2dechoes}). The same time constant is observed on
the $\rm{P_{b0}}$ echo. The transients as well as the echoes and
their decay are independent of temperature up to 12~K. Above 12~K,
no echo is observed due to the decrease in the amplitudes of the
underlying current transient.

At least two physical processes can contribute to the decay of the
echo, namely (i) the actual decoherence usually denoted by the
transverse relaxation time $T_2$ and (ii) the loss of spin-pairs
through recombination. If the two processes are independent,
Matthiessen's rule predicts that the faster process will govern the
overall echo decay.

Let us first assume that the echo decay of caused by fast $T_2$
processes. Comparing the observed echo decay time $\tau_{echo}$ with
reports on decoherence in Si:P from the literature, it becomes clear
that the value reported here is significantly shorter than in
earlier experiments via conventional pulsed ESR to values between
$100~\rm{\mu s}$ and 60~ms \cite{tyryshkin03, gordon58}. Those
experiments are typically performed at low doping concentrations
$(<2\times 10^{16}~\rm{cm^{-3}})$ and in bulk samples to ensure the
isolation of the phosphorus spins.  Chiba and Hirai \cite{chiba72}
studied the decoherence time as function of the phosphorus doping
concentration at 1.5~K and observed a significant decrease of $T_2$
by two orders of magnitude from $300~\rm{\mu s}$ to $5~\rm{\mu s}$
at a doping concentration of $1-2\times 10^{17}~\rm{cm^{-3}}$
similar to the doping concentration used in this experiment.
Therefore, the $\tau_{echo}$ of $1.7\pm0.2~\rm{\mu s}$ determined
here could already be accounted for by the high doping
concentration.

Additionally, in this experiment the P donors are particularly close
to the Si/SiO$_2$ interface expected to be necessary for the tuning
of the hyperfine interaction and of the exchange between neighboring
donors by gate potentials when using P in devices for quantum logic
\cite{kane98, martins04}. De~Sousa \cite{desousa07} has calculated
the influence of magnetic noise caused by paramagnetic
Si/SiO$_2$-interface states on the decoherence times of donors and
estimates $T_2=4\times 10^{-8}~{\rm{s}}\times(d/{\rm{nm}})^2$ for an
interface spin density of $10^{11}~\rm{cm^{-2}}$, where $d$ denotes
the distance of the donors to the interface. For these interface
state densities, Schenkel et al.~\cite{schenkel06} have determined
the decoherence of Sb donors in $^{28}$Si by conventional pulsed ESR
for Sb implanted to comparatively large depths of 50 and 150~nm, in
good agreement with the theory. Taking into account the considerably
smaller distance to the interface studied here and the slightly
larger $\rm{P_{b0}}$ density of $10^{12}~\rm{cm^{-2}}$ of native
oxides \cite{pierreux02} our data can similarly be accounted for
well by de~Sousa's prediction \cite{desousa07}. Furthermore, the
temperature independence of $\tau_{echo}$ suggests that spin
flip-flop processes between the P and the $\rm{P_{b0}}$ spins.

Alternatively, the echo decay could also be due to recombination.
Rate limiting for Carr-Purcell echoes would be the singlet
recombination rate $r_S$ \cite{boehme03}. Since the singlet content
of the spin pair during the evolution periods of our echo experiment
is $1/4$, $\tau_{echo}=4/r_S$ in the case of slow $T_2$ processes.
For P-$\rm{P_{b0}}$ spin pairs, the singlet recombination time has
not been determined independently. However, effective lifetimes of
photo-generated charge carriers can be as short as $1~\rm{\mu s}$,
as has been studied for P-doped emitters in crystalline Si solar
cells \cite{king90}, structures very similar to our test device. The
value of $\tau_{echo}$ found would in this case correspond to
$r_S=2.3\times10^6~\rm{1/s}$, a rate significantly higher than
observable directly from the current transients. Since EDMR
experiments on the identical structures using the inversion recovery
pulse sequence are limited by the same $r_S$ \cite{hoehne07unpub},
we conclude that, at the densities of P and $\rm{P_{b0}}$ in our
sample, indeed recombination determines the echo decay. Furthermore,
the inversion recovery EDMR results suggest that microwave phase
noise can be excluded as the origin of the echo decay.

As discussed, the P-$\rm{P_{b0}}$ pair interaction allows the purely
electrical detection of the spin states of donor qubits.
Irrespective of the actual microscopic mechanism, the echo decay
time determined by pulsed EDMR is the relevant timescale on which
such pairs can be used for quantum information processing. The
$\tau_{echo}$ found here allows to apply around 30 $\pi$-pulses of
60~ns length. While this is below the standard DiVincenzo
requirement, this should be sufficient to demonstrate the basic
useability of P donors in quantum information processing. Still, a
variation of the phosphorus density, the density of the interface
states, the isotopic composition of the host lattice and the
illumination intensity should allow to reduce the decoherence
processes and increase the recombination times in such structures
extending the time scale on which P-$\rm{P_{b0}}$ pairs can be used
for quantum logic.

In conclusion, we have demonstrated the electrical detection of
Carr-Purcell echoes in 15~nm thin Si:P structures. The decay of the
echo amplitude is mono-exponential with a decay time constant of
$1.7\pm0.2~\rm{\mu s}$. Several mechanisms such as the relatively
high phosphorus doping concentration, the small P/SiO$_2$-interface
distance, and the illumination of the sample will contribute to the
overall decoherence process. The systematic determination of the
echo decay time as a function of spin density, interface distance
and illumination will provide further experimental data on the
dynamics of the P-$\rm{P_{b0}}$ spin pairs, which are highly
relevant for the use of this system for the read-out of donor spins.
However, the results also directly address other donor-based quantum
computer architectures with different spin read-out techniques,
since $\rm{P_{b0}}$ paramagnetic centers are present at all
Si/SiO$_2$ interfaces. Furthermore, the determination of decoherence
via Carr-Purcell echoes will allow a more detailed understanding of
the complex charge carrier and spin dynamics in pulsed EDMR.
Finally, the combination of echo tomography and pulsed electrically
detected magnetic resonance opens the possibility to apply further
pulse sequences including free evolution times to study spin-spin
interactions in Si with unprecedented sensitivity.

The authors would like to thank G\"{u}nther Vogg and Frank Bensch for
sample growth and Christoph Scheurer and Frank H. L. Koppens for
valuable discussions. This work was supported by the Deutsche
Forschungsgemeinschaft through SFB~631.

\bibliography{HHmod2}% Produces the bibliography via BibTeX.
\clearpage
\end{document}